\documentclass[aps,prd,reprint,amsmath,amssymb,showpacs,superscriptaddress]{revtex4-1}

\usepackage{graphicx}
\usepackage{subfigure}
\usepackage{dcolumn}
\usepackage{bm}
\usepackage{slashed}
\usepackage{ulem}
\usepackage{MnSymbol}
\usepackage{epstopdf}
\usepackage{mathrsfs}
\usepackage{color,xcolor}

\usepackage{hyperref}

\begin{document}

\title{ Mass-dependence of pseudocritical temperature in mean field approximation}

\author{Zhan Bai}
\affiliation{Department of Physics and State Key Laboratory of Nuclear Physics and Technology, Peking University, Beijing 100871, China.}

\author{Lei Chang}
\email{Corresponding author. changlei@nankai.edu.cn}
\affiliation{School of Physics, Nankai University, Tianjin 300071, China}

\author{Jingyi Chao}
\affiliation{Institute of Modern Physics, Chinese Academy of Sciences, Lanzhou, 730000, China}

\author{Fei Gao}
\affiliation{Institut f{\"u}r Theoretische Physik,
	Universit{\"a}t Heidelberg, Philosophenweg 16,
	69120 Heidelberg, Germany }

\author{Yu-Xin Liu}
\email{Corresponding author. yxliu@pku.edu.cn}
\affiliation{Department of Physics and State Key Laboratory of Nuclear Physics and Technology, Peking University, Beijing 100871, China.}
\affiliation{Collaborative Innovation Center of Quantum Matter, Beijing 100871, China.}
\affiliation{Center for High Energy Physics, Peking University, Beijing 100871, China.}


\date{\today}

\begin{abstract}
We implement the Dyson-Schwinger equations approach to study the mass dependence of pseudocritical temperature of QCD phase transition at zero chemical potential.
We restrict our computation in the mean field approximation which could lead to a clear critical behavior.
We analyze the scaling behavior with different shape of interaction kernel by considering different dressed-gluon models.
The critical exponent we obtained is consistent with that in the $3D$ $\textrm{O}(4)$ universality class.
The size of critical region is up to $m_{0}^{} \le 2\sim 4\;$MeV in this mean field approximation which sets naturally an upper bound of the critical region
since the fluctuations beyond mean-field  usually diminish the critical region.
Besides, we analyze the possible percentage of the maximum chiral susceptibility and pion mass range at which the chiral phase transition temperature is independent of
the current quark mass.
The results show that the percentage and the pion mass range depend on the details of interaction kernel, which differs in gluon models.
\end{abstract}

\maketitle

\section{Introduction}
\label{sec:Introduction}

The basic degrees of freedom in Quantum Chromodynamics (QCD)$-$ gluons and quarks $-$ do not exist as asymptotic states, {\it i.e.}, these partonic excitations
do not propagate with integrity over the length-scales that exceed some modest fraction of the proton's radius.
%
%
The forces responsible for this phenomenon also appear to generate more than $98\%$ of the mass of visible matter.
These features are  known as confinement and dynamical chiral symmetry breaking (DCSB). Theoretically,
these features could be described  via the appearance of momentum dependent
mass-functions for quarks  and gluons even in the absence of any Higgs-like mechanism~\cite{Bhagwat:2003,Yuan:2006,Chang:2007,Braun:2008pi,Braun:2009gm,Fister:2011uw,Mitter:2014wpa,Braun:2014ata, Rennecke:2015eba, Fu:2016tey, Rennecke:2016tkm, Cyrol:2016tym, Cyrol:2017ewj, Cyrol:2017ewj, Cyrol:2017qkl, Fu:2018qsk, Fu:2019hdw, Leonhardt:2019fua, Braun:2019aow, Roberts:2000aa, Qin:2010nq, Fischer:2011mz, Fischer:2013eca, Fischer:2014ata, Eichmann:2015kfa,  Gao:2016hks,Gao:2017gvf, Gao:2020qsj,Fischer:2018sdj, Gunkel:2019xnh, Isserstedt:2019pgx, Reinosa:2015oua, Reinosa:2016iml, Maelger:2017amh, Maelger:2018vow, Maelger:2019cbk, Aguilar:2016lbe, Aguilar:2017dco, Aguilar:2018epe,Bazavov:2012vg, Borsanyi:2013hza, Borsanyi:2014ewa, Bonati:2015bha, Bellwied:2015rza, Bazavov:2017dus, Bazavov:2017tot, Bonati:2018nut, Borsanyi:2018grb, Bazavov:2018mes, Guenther:2018flo, Ding:2019prx}.
While including the medium effect at finite temperature and chemical potential, these features will lead the QCD matter to have a rich phase structure as the thermal fluctuation dramatically affect the momentum dependence of mass kernel.
Considering the chiral symmetry property,
as the temperature and chemical potential increases, owing to the asymptotic freedom of QCD~\cite{Politzer:2005kc,Wilczek:2005az,Gross:1998bd}, QCD matter transits from a DCSB  phase to a chiral symmetry preserving (CS)  phase.
Such features cover a vast array of empirical aspects, from gluon and quark interactions at the highest energies achievable with the heavy ion collision, to the nature of nuclear matter in the inner part of a compact star, which then deliver a sketch of QCD phases at finite temperature and chemical potential~\cite{Luo:2017faz,Adamczyk:2017iwn,Andronic:2017pug,Stephanov:2007fk,Andersen:2014xxa,Shuryak:2014zxa,Pawlowski:2014aha,Roberts:2000aa,Fischer:2018sdj,Yin:2018ejt}.

It has been shown that if one sets the up and down quarks in chiral limit, the crossover behavior of the matter at zero chemical potential steepens into 
a second order chiral phase transition  which belongs to the 3-dimensional ($3D$) $\textrm{O}(4)$ universality class,
if the $U_{A}^{}(1)$ anomaly still exists in the phase transition region~\cite{Grahl:2013pba,Sato:2014axa,Ding:2019prx,Braun:2020ada,Gao:2015kea}.
If $U_{A}^{}(1)$ symmetry has been restored sufficiently, it is expected to be a larger $3D$ universality class.
At physical mass, the critical end point (CEP) is also expected to be $3D$ Ising class~\cite{Gao:2015kea,Hatta:2002sj}.
Based on these critical behavior analysis,  the phase structure at vanishing density could be helpful for determining  the CEP at the physical quark mass,
for instance, the phase transition temperature at chiral limit could be considered as the upper bound  for the temperature of CEP.

However, the reconstruction with a certain universality class is subtle owing to the uncertainty of the size of the critical region.
Generally, the fluctuations drive deviation of the scaling behavior,
and a detailed analysis reveals that the actual size of the scaling behavior could be only up to $m_{\pi}\sim 1\;$MeV~\cite{Klein:2017shl}.

Since the Dyson-Schwinger equations (DSEs) approach has been shown to be a sophisticated QCD approach~\cite{Roberts:2000aa,Review1,Review2,Review3},
we study the scaling behavior of chiral susceptibility in the DSEs approach in this paper.
We will focus on the computation in rainbow approximation which is generally working at mean-field approximation level~\cite{Hoell:1999prc}.
Under the mean field approximation, it would have clearer signal of critical behavior and broader critical region.
Besides, we take three different dressed-gluon models to explore the effect of the interaction feature on the critical behavior.

The remainders of this article are organized as follows.
In Sec.~\ref{Framework} we reiterate briefly the DSEs approach at finite temperature
including the models and the chiral phase transition criterion.
Sec.~\ref{Results} represents our results of the chiral susceptibility and the pseudocritical temperature, as well as their current mass dependence.
Finally, we summarize in Sec.~\ref{Summary}.

\section{Quark gap equation  at finite temperature}
\label{Framework}

The quark propagator at finite temperature can be determined by the gap equation
\begin{eqnarray}
\label{eq:gap1}
S(\vec{p}, \omega_n)^{-1} \! &=& \! Z_{2}i\vec{\gamma}\cdot\vec{p}
+ \! Z_{2C}i\gamma_4   \omega_{n} \! + \! Z_{4} m_{0}^{} \! + \! Z_{1} \Sigma(\vec{p},  \omega_{n}) \, , \quad \\
\nonumber
\Sigma(\vec{p}, \omega_n) &=& T\sum_{l=-\infty}^\infty \! \int\frac{d^3{q}}{(2\pi)^3}\; {g^{2}} D_{\mu\nu} (\vec{p}-\vec{q}, \Omega_{nl}; T, \mu)\\
& & \times \frac{\lambda^a}{2} {\gamma_{\mu}} S(\vec{q},
 \omega_{l}) \frac{\lambda^a}{2}
\Gamma_{\nu} (\vec{q},  \omega_{l},\vec{p}, \omega_{n})\, ,
\label{eq:gap2}
\end{eqnarray}
where $Z_{1},\, Z_{2},\, Z_{2C}, Z_{4}$ are the renormalization constants, $m_{0}^{}$ is the current quark mass,  $\omega_n=(2n+1)\pi T$ being the quark Matsubara frequency ,
$\Omega_{nl} = \omega_{n} - \omega_{l}$; $D_{\mu\nu}$ is the dressed-gluon propagator; and $\Gamma_{\nu}$ is the dressed-quark-gluon interaction vertex.
%
%

According to the Lorentz structure analysis, the gap equation's solution can be decomposed as
\begin{eqnarray}\label{eq:qdirac}
\nonumber
S(\vec{p}, \omega_n)^{-1} & = & i\vec{\gamma} \cdot \vec{p}\, A(\vec{p}\,^2,  \omega_{n}^2) \\
&& + i\gamma_{4}  \omega_{n} C(\vec{p}\,^2,  \omega_{n}^2) + B(\vec{p}\,^2,  \omega_{n}^2) \, .
\end{eqnarray}
%

In our calculation, we take the approximation  $\Gamma_{\nu}(\vec{q}, \tilde\omega_{l},\vec{p},\tilde\omega_{n}) = \gamma_{\nu}\,$ for the quark-gluon interaction vertex
which is referred as the rainbow approximation.
The rainbow approximation is the leading order approximation~\cite{Munczek:1995prd,Bender:1996plb}. It is essentially a mean-field approximation~\cite{Hoell:1999prc}, which will then lead to a clear analysis of the universality class.

To solve the gap equation, we should also have the information of the dressed-gluon propagator. The dressed-gluon propagator is usually approximated as
\begin{equation}
g^{2} D_{\mu\nu}(\vec{k}, \Omega_{nl}) = \mathcal{G}(k^{2})D_{\mu\nu}^{\textrm{free}}(k)\,,
\end{equation}
where $D_{\mu\nu}^{\textrm{free}}$ is the free gluon propagator
\begin{equation}
D_{\mu\nu}^{\textrm{free}}(k)= \left(\delta_{\mu\nu}-\frac{k_{\mu}k_{\nu}}{k^{2}}\right)\frac{1}{k^{2}},
\end{equation}
and $k_{\mu}=\left(\vec{k},\Omega_{nl}\right)$.
The $\mathcal{G}(k^{2})$ is the effective interaction which depends on models.

Recent studies of QCD's gauge sector confirmed a massive gluon propagator on the domain at $q^{2}=0$~~\cite{Bowman:2004jm,Cucchieri:2007ta,Boucaud:2010gr,Oliveira:2010xc,Fister:2011uw,Cyrol:2016tym, Cyrol:2017qkl,Gao:2017tkg, Aguilar:2008xm,Aguilar:2012rz,Aguilar:2015nqa,Aguilar:2019kxz,Aguilar:2019uob}.
The behavior could be well described by the infrared constant model, known as the Qin-Chang (QC) model~\cite{Qin:2011dd}.
The QC gluon model is a  QCD based model with the correct momentum dependence of QCD, which reads
\begin{eqnarray}
\frac{\mathcal{G}({k^{2}})}{k^{2}} & = & 8{\pi^{2}} D
\frac{1}{\omega^{4}} e^{-{k^{2}}/\omega^{2}} \notag\\
& & + \frac{8{\pi^{2}} {\gamma_{m}}}{{\ln}[ \tau \! + \! (1 \! + \!
{k^{2}}/{\Lambda_{\text{QCD}}^{2}} ) ^{2} ] } \,
{\cal F}(k^{2}) \, ,
\end{eqnarray}
with ${\cal F}(k^{2}) = (1-\exp(-k^{2}/4 m_{t}^{2})/k^{2}$, $\tau=e^2-1$, $m_t=0.5\,$~GeV, $\gamma_m=12/25$, and $\Lambda^{}_{\text{QCD}}=0.234$~GeV.

There are two parameters, $D$ and $\omega$ in the QC model.
In vacuum, we can take $(D\omega)^{1/3}= 0.82 \, \mathrm{GeV}$ and $\omega = 0.5\,$GeV together with light quark mass  $m_{0}^{\zeta=2~\textrm{GeV}}=6.6$~MeV,
which reproduces pion properties $m_{\pi}^{} = 0.14\,$GeV and $f_{\pi} = 0.092\,$GeV~\cite{Chen:2018rwz}.

Noticing that the function $\mathcal{G}(k^2)$ in above equation is served as a momentum distribution of interaction,
this model could cover various types of the interaction via varying the $\omega$ which stands for the interaction width of QCD.
Except for the perturbative correction, as $\omega$ goes to 0, the distribution becomes a $\delta$-function, which is just the so called Munczek-Nemirovsky (MN) model~\cite{MN-Model,Maris:1997eg}.
For very large $\omega$, it becomes approximately the contact model
%
%
which is then a representative of NJL mdoel in DSEs approach~\cite{Roberts:2011wy}.

At finite temperature, the MN model could be generalized as
\begin{equation}\label{eq:MunczekGluon}
g^2 D_{\mu\nu}(\vec{p},\Omega_{k})=\left(\delta_{\mu\nu}-\frac{p_{\mu}p_{\nu}}{|\vec{p}|^{2}+\Omega_{k}^2}\right) 2\pi^{3}\frac{\eta^{2}}{T}\delta_{k0}\delta^{3}(\vec{p}),
\end{equation}
where $\Omega_{k}=2k\pi T$ is the boson Matsubara frequency and $(p_{\mu})=(\vec{p},\Omega_{k})$. $\eta$ is a mass-scale parameter.
Ref.~\cite{Maris:1997eg} shows that with parameters $\eta=1.37\;$GeV and $m=30\;$MeV, one can get the pion mass $m_{\pi}=140$MeV.

The contact model reads
\begin{equation}
g^{2}D_{\mu\nu}=\delta_{\mu\nu}\frac{1}{m_{G}^{2}},
\end{equation}
where $m_{G}$ is an interaction strength parameter and it can be taken $m_{G}=0.132$ GeV in vacuum~\cite{Roberts:2011wy}.

At finite temperature, however, the interaction strength should be modified.
For the QC model, we multiply a damping factor to the coupling constant D:
\begin{equation}\label{eqn:damping}
D(T)=D\times \frac{1}{\left(1+\frac{\alpha T}{m_{\pi}}\right)^{2}}.
\end{equation}
And for the MN model and contact model, the same damping factor $1/(1+\alpha T/m_{\pi})^{2}$ can be directly multiplied to the dressed-gluon propagator.

The dimensionless damping parameter $\alpha$ is calibrated to reproduce the critical temperature $T_{c}=156.5\;$MeV at physical pion mass~\cite{Braun:2020ada}.
One has then $\alpha=0.1731$, $0.5274$ and $0.1477$ for the QC model, MN model and contact model, respectively.

The chiral symmetry property manifests by the quark propagator straightforwardly.
The order parameter of chiral symmetry is usually defined as the quark condensate which is an integral of the quark propagator.
This definition needs an ultraviolet subtraction at finite current quark mass(see, e.g. Ref~\cite{Gao:2016qkh} and the reference therein).
Besides, noticing that the dominant contribution in the ultraviolet region is perturbative due to the asymptotic free behavior,
and the DCSB effect is generated mainly by the  infrared domain,
the order parameter could also be interpreted as the running mass of the quark at zero momentum.
The different definitions of the order parameter will not change the phase diagram for the first and second order phase transitions.
It changes slightly the location of the pseudocritical temperature in crossover region,
but the critical exponents are universal and will not be altered.
Therefore, in this paper, we would simply take $B(\omega_{0},\vec{p}=0)$ as the order parameter.

After then, the chiral susceptibility could be defined as the derivative respective to the current quark mass~\cite{Qin:2010nq,Ding:2019prx,Braun:2020ada,Gao:2016qkh}:
\begin{equation}
\chi=\frac{\partial B(\omega_{0},\vec{p}=0)}{\partial m_{0}} \, .
\end{equation}
The pseudocritical temperature is then just the temperature for the $\chi$ to reach its maximum.
Such a definition of the chiral susceptibility is not exactly the same as that used in lattice QCD and functional renormalization group calculations,
however the difference between the extracted pseudocritical temperature should be quite limited~\cite{Gao:2016qkh}.
Therefore the exact value of chiral susceptibility in this paper is different from those in Refs.~\cite{Ding:2019prx,Braun:2020ada},
but the obtained pseudocritical temperatures are comparable (see next Section).

\section{Results and Discussions}
\label{Results}

\begin{figure}[t]
\center
	\includegraphics[width=0.45\textwidth]{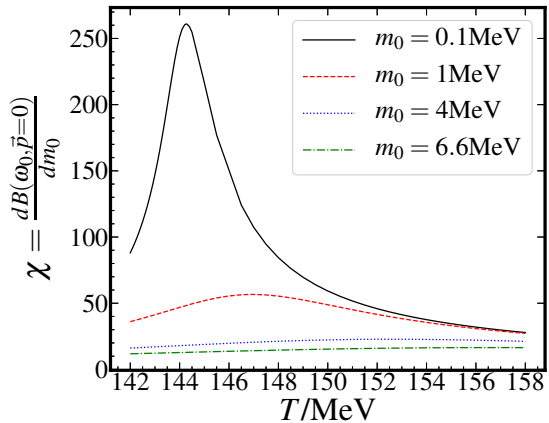}
	\caption{(color online) Calculated chiral susceptibility as a function of temperature for several current quark masses.}
\label{fig:ChiralSusceptibility}
\end{figure}

In this section, we present our numerical results calculated by the DSE with the three models mentioned above.
In Fig.~\ref{fig:ChiralSusceptibility}, we show the chiral susceptibility as a function of temperature for several values of the current quark mass $m_{0}$ in the QC model.
It is evident that the peak gets narrower and narrower as the current quark mass becomes very small.
At chiral limit, the chiral susceptibility becomes divergent which means it becomes the second order phase transition~\cite{Qin:2010nq}.
The critical temperature at chiral limit is $T_{c}=143.6$MeV.
Similar behaviors could also be found in the other two models, and we obtain $T_{c}=142.9\;$MeV for the contact model, and $T_{c}=142.0\;$MeV for the MN model in chiral limit.

\begin{figure}[t]
\center
	\includegraphics[width=0.45\textwidth]{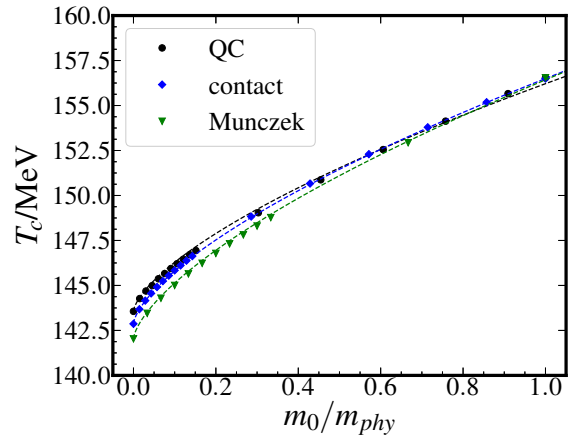}
	\caption{(color online) Calculated critical temperature as a function of current quark mass rescaled by  $m_{\textrm{phys}}$ which is the quark mass
        corresponding to the physical pion mass $m_{\pi}=140\;$MeV.
		  The different symbols with different colors correspond to the result calculated using different model,
	The dashed lines correspond to the fit of the calculated data using the formula $T=a\times m_{0}^{2/3}+T_{c}^{\textrm{chiral}}$.}
\label{fig:m0_Tc}
\end{figure}

In Fig~\ref{fig:m0_Tc}, we present the calculated result of the pseudocritical temperature as a function of the current quark mass.
The current quark mass corresponding to the physical pion mass is $m_{0}=6.6\;$MeV for the QC model~\cite{Chen:2018rwz},
$m_{0}=7\;$MeV for the contact~\cite{Roberts:2011wy}, and $30\;$MeV for the MN model~\cite{Maris:1997eg}.
The pseudocritical temperature for all the three models have been tuned to $156.5\;$MeV by altering $\alpha$ in Eq.~(\ref{eqn:damping}).
Although the three models distinct from each other greatly,
the obtained pseudocritical temperature is nearly the same for different models.
This means that the pseudocritical temperature relies on the model only slightly.

In order to further study the behavior of the phase transition,
we take a formula $T=a m_{0}^{b}+c$ to fit the data.
We fix $c = T_{c}^{\textrm{chiral}}$ with $T_{c}^{\textrm{chiral}}$  the critical temperature at $m_{0}=0$ since there's no ambiguity in our chiral limit results as the susceptibility becomes divergent.
For the power index $b$,
we could make use of the value of the critical exponent as $b=1/\beta\delta=2/3$ in our mean field approximation, which is consistent with the $3D$ $\textrm{O}(4)$ universality class~\cite{Hoell:1999prc,Hatta:2002sj}.
As can be seen from the figure, the calculated result can be fitted quite well with the mean field critical exponent.
Such a result is also consistent with that given by analyzing the heat capability in the DSE approach of QCD~\cite{Gao:2015kea}

\begin{figure}[t]
\center
	\includegraphics[width=0.45\textwidth]{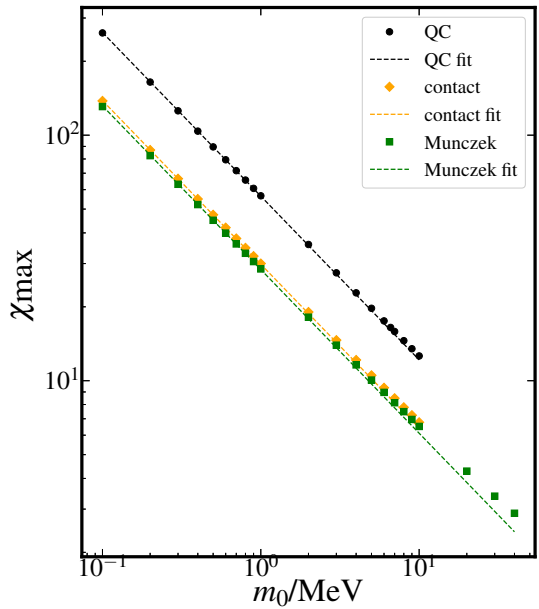}
	\caption{(color online)Calculated chiral susceptibility as a function of current quark mass.
	The dashed lines represent the fitted results of the calculated data using the formula $\chi=A m^{ 1-1/\delta}$ with $\delta=3$.}
\label{fig:chimax_m0}
\end{figure}

Moreover, we compute the scaling behavior of chiral susceptibility with respect to the current quark mass in the QC model.
The universality class analysis gives that~\cite{Hoell:1999prc}:
\begin{equation}\label{eqn:univer_class}
\chi=A m_{0}^{1-1/\delta}.
\end{equation}
In Fig.~\ref{fig:chimax_m0}, we illustrate the calculated data and the fitted result.
It is apparent that we can assign the parameter as $\delta=3$, which is consistent with $3D$ $\textrm{O}(4)$ universality class.
In some details, we can extract that the critical region is up to $m_{q} \sim 4\;$MeV for the QC model,
and for the contact and the MN model, the critical region is $m_{0}\lesssim 2\;$MeV.
This means that the MN model has a smaller critical region than the contact model,
since the physical current quark mass is larger in the MN model.
After analyzing the scaling behavior of the chiral susceptibility with respect to temperature, we get another critical exponent  as $\beta=1/2$.

The scaling behavior could also be demonstrated in terms of the mass of pion since the mass of pion is an experimental quantity,
therefore, it is better than the current quark mass which is dependent on the renormalization schemes.
It has been well known that the Gell-Mann--Oakes--Renner (GOR) relation indicates
a simple connection between the pion mass and the current quark mass near the chiral limit, which reads
 \begin{eqnarray}\label{eq:gmor}
f^{2}_{\pi} m^{2}_{\pi} = - 2m(\zeta)\langle\bar{q}q\rangle\,,
\end{eqnarray}
with $f_{\pi}^{}$ the pion decay constant, $m(\zeta)$ is the current quark mass at renormalization scale $\zeta$.
In this paper, we can then take the leading order of this relation and have
\begin{equation}
m_{\pi} = C\sqrt{m_{0}} \, ,\, \textrm{in the unit of MeV},
\end{equation}
where $C$ is a coefficient depending on the model.
The value of $C$ is determined by fitting the results from Bethe-Salpeter equation calculation,
and is $54.49$ for QC model~\cite{Chen:2018rwz},
$52.76$ for contact model~\cite{Roberts:2011wy},
and $25.56$ for MN model~\cite{Maris:1997eg}.
It is clear that the pion mass obtained via the DSE with the RL truncation and the QC gluon model is quite the same with that given in the contact model,
while differs from that in the MN model since the $m_{0}^{}$ in the MN model is much larger than that in the other models.

\begin{figure}[t]
\center
	\includegraphics[width=0.45\textwidth]{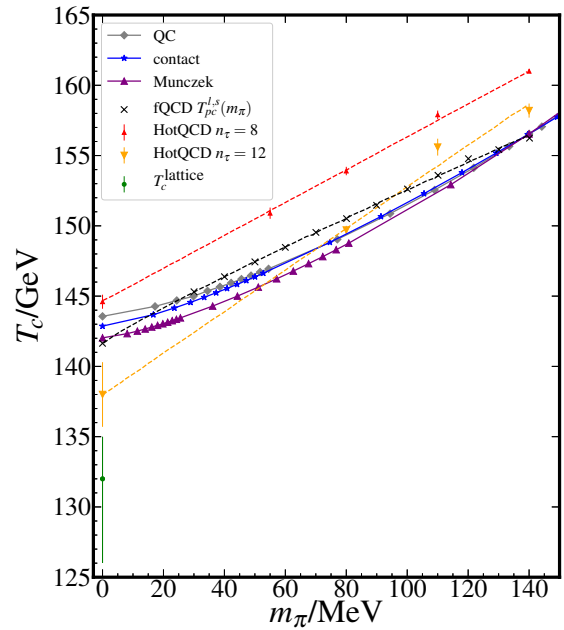}
	\caption{(color online) Calculated pseudocritical temperature as a function of pion mass.
		The fQCD result is taken from Ref.~\cite{Braun:2020ada}, and the HotQCD result from Ref.~\cite{Ding:2019prx}.
	 }
\label{fig:mpi_Tc}
\end{figure}

The calculated result of the $T_{c}$ as a function of $m_{\pi}$ is shown in Fig.~\ref{fig:mpi_Tc}.
For comparison we plot also the result obtained with the functional renormalization group approach of QCD (fQCD)~\cite{Braun:2020ada}
and that given in lattice QCD
(HotQCD)~\cite{Braun:2020ada,Ding:2019prx}.
One can notice from Fig.~\ref{fig:mpi_Tc} that,
in contrast to the linear dependence from fQCD and HotQCD result,
our calculated $T_{c}$ has a $\sim 4/3$ power dependence on the pion mass.
Beyond the critical region, we can infer from the figure that the relation between $T_{c}$ and $m_{\pi}$ goes gradually to a simple linear relation.
Recalling that  the critical exponent of 3$D$ $\textrm{O}(4)$ class beyond mean-field is around $1$~\cite{Ding:2019prx,Braun:2020ada},
therefore, the full QCD computation could find a linear relation between $T_c$ and $m_\pi$ without  the limitation of critical region.

\begin{figure}[t]
\center
	\includegraphics[width=0.45\textwidth]{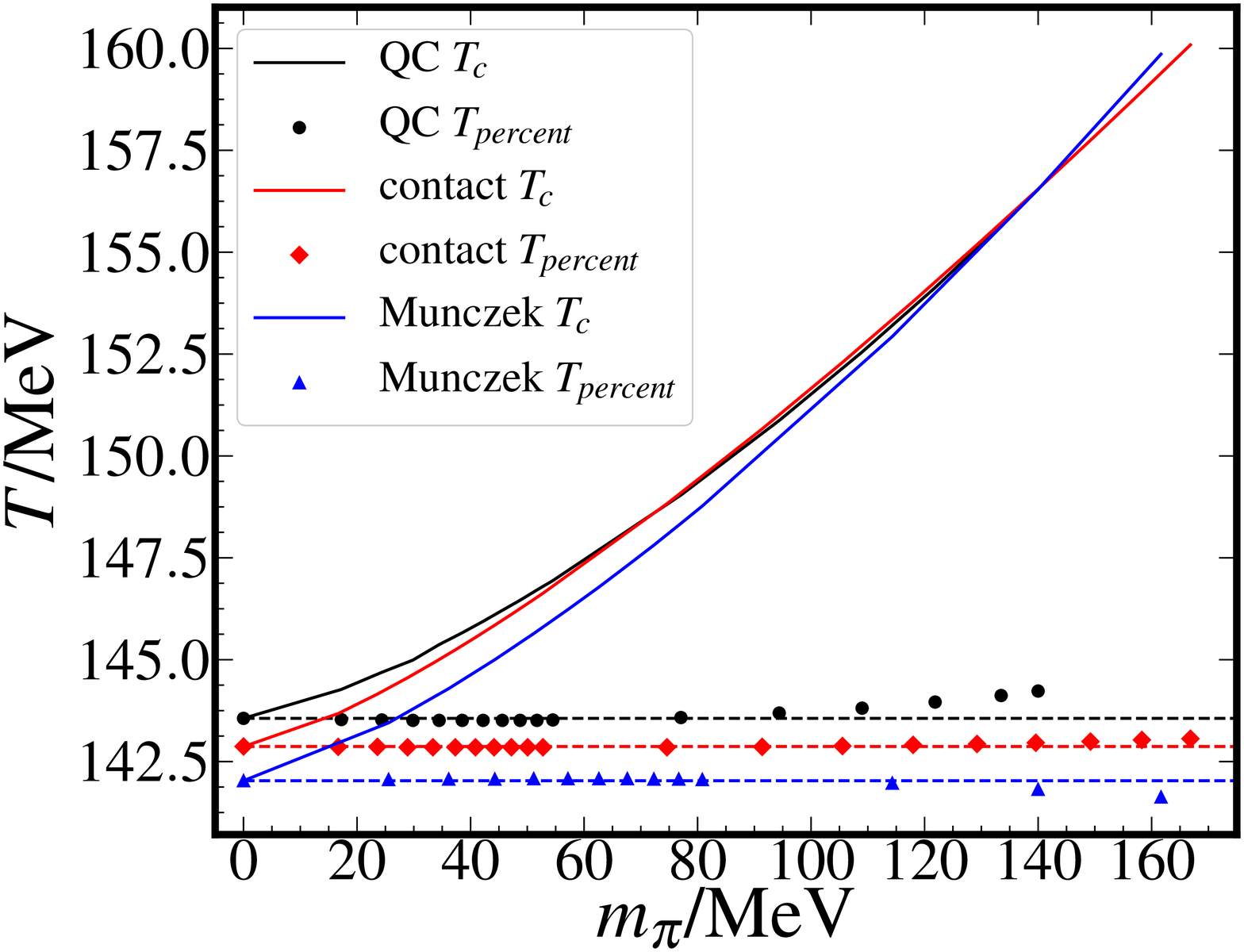}
	\caption{(color online) Comparison of the pion mass dependence of the typical pseudocritical temperature ($T_{c}^{}$),
        the extrapolated pseudocritical temperature ($T_{\textrm{percent}}$), and the calculated critical temperature in chiral limit, in the three models.
	      The solid lines correspond to $T_{c}$s calculated in the three models.
		    The symbols {\small $\bullet$}, {\color{red}{\small $\blacklozenge$}} and {\color{blue}{\small $\blacktriangle$}} stand for the $T_{\textrm{percent}}$ with the percentage being $78\%$, $79\%$, $80\%$ for the QC,
            the contact, and the MN model, respectively.
        	  The horizontal dashed lines represent the $T_{\textrm{chiral}}$ in the three models.
	 }
\label{fig:Tc_ratio}
\end{figure}

In the lattice QCD simulation, it has been argued that the temperature at which the chiral susceptibility is $60 \%$ of its maximum
remains nearly the same for a large range of quark mass, so that the critical temperature for the chiral symmetry to be restored in chiral limit is obtained by extrapolation~\cite{Ding:2019prx}.
In order to investigate such an extrapolation scheme, we present the calculated critical temperature with a certain percentage of the maximal susceptibility
via the three models in Fig.~\ref{fig:Tc_ratio}.
The solid lines in Fig.~\ref{fig:Tc_ratio} correspond to a percentage of $100 \%$, {\it i.e.}, they correspond to the typical pseudocritical temperature, $T_{c}$.
The horizontal dashed lines stand for the critical temperature in the chiral limit, $T_{\textrm{chiral}}$.
The symbols {\small $\bullet$}, {\color{red}{\small $\blacklozenge$}} and {\color{blue}{\small $\blacktriangle$}} represent the temperature corresponding to a percentage of the maximal chiral susceptibility, $T_{\textrm{percent}}$, in the three models respectively.
%

As we can see from the Fig.~\ref{fig:Tc_ratio}, for all the three models, when the pion mass (and the current quark mass) is large enough,
the $T_{\textrm{percent}}$ deviates from the horizontal dashed line, $T_{\textrm{chiral}}$, in the corresponding model.
This means that when we take a fixed percentage of the $\chi_{\textrm{max}}$ as a criterion,
the pseudocritical temperature is not invariable for a large range of pion mass.

However, if we survey the obtained results with various percentages carefully, we observe that there exist a pion mass range where $T_{\textrm{percent}}$ remains almost invariable.
And the percentage for the pion mass range to reach its maximum is $78\%$, $79\%$ and $80\%$ for the QC model, the contact model and the MN model, respectively.
As we can see from the figure, for the contact model and the MN model, the $T_{\textrm{percent}}$ remains unchanged up to physical pion mass,
but the $T_{\textrm{percent}}$ begins to increase at $m_{\pi}\sim 100$MeV in the QC model.


\section{Summary}
\label{Summary}

It has been argued that the phase transition of QCD at chiral limit belongs to the $3D$ $\textrm{O}(4)$ universality class.
The universality class analysis gives the critical  behavior at the chiral limit as that relation between the pseudocritical temperature
and the current quark mass behaves as $T_{c} \sim m_{q}^{ 1/\beta\delta}$.
We restrict our computation in the mean field approximation of QCD which could lead to a clear critical behavior, and give $1/\beta\delta=2/3$,
which is consistent with the universality class result.
We analyze the scaling behavior with different features of interaction by considering different dressed-gluon models.
The obtained critical exponent is consistent with the $3D$ $\textrm{O}(4)$ universality class.
The size of critical region is up to $m_{0}\sim 2\sim 4$ MeV in this mean field approximation which sets naturally an upper bound of the critical region
since the fluctuations beyond mean-field rely usually on a more subtle universal classification of the system and blur the regime of critical behaviors.

We have implemented three different types of dressed-gluon models which represent different momentum dependent behaviors of the interaction kernel.
Our obtained results lead us to a conclusion that the momentum dependence of the interaction kernel does not change the mean field approximation.
To go beyond the mean field approximation, people need to either introduce the complicated tensor structures of the quark-gluon interaction vertex
of which the coefficients involve explicitly the quark mass dependence~\cite{Tang:2019} or employ the effective pion exchange which also contains the mass dependence via the pion mass.

Another interesting quantity has been marked by lattice QCD simulation analysis is the pseudocritical temperature where the chiral susceptibility is the $60\%$ of the maximum.
At this point, the temperature is found to be independent of current quark mass.
However, our results indicate that this is only true when the current quark mass and pion mass is not too large.
For the QC model, this temperature is invariant only for $m_{\pi}\lesssim 100\:$MeV,
while for the contact and the MN model it can be invariant up to the physical pion mass.
The fixed percentage of the maximal chiral susceptibility is also different for different models,
which reads $78\%$, $79\%$ and $80\%$ for the QC, the contact and the MN model.

Despite of the mean field approximation, here we obtain the chiral phase transition temperature at chiral limit as $T_c^0=142.8\pm 0.8$MeV with the error from the different gluon models.

\section{Acknowledgement}

The work was supported by the National Natural Science Foundation of China under Contracts No. 11435001, No. 11775041, No. 11947108, and No. 11605254.
FG is grateful for the support from Alexander von Humbodlt Foundation.

\end{document}